\documentclass[sigconf]{acmart}

\usepackage{booktabs} % For formal tables
\usepackage{upgreek} % For uppercase greek
\usepackage{tikz-cd}
\usepackage{tikz}
\usetikzlibrary{shapes,arrows}
\usepackage{hyperref}
\usepackage{graphicx,verbatimbox}
\usepackage{listings}
\usepackage{subcaption}
\usepackage{mathtools}
\usepackage{color}
\usepackage{tabularx,ragged2e}
\usepackage{multirow}
\usepackage{mathtools}
\usepackage{enumitem}
\usepackage[british]{babel}
\usepackage{xspace} 
\usepackage{listings}

\colorlet{punct}{red!60!black}
\definecolor{background}{HTML}{EEEEEE}
\definecolor{delim}{RGB}{20,105,176}
\colorlet{numb}{magenta!60!black}

\lstdefinelanguage{json}{
	basicstyle=\normalfont\ttfamily,
	numbers=left,
	numberstyle=\scriptsize,
	stepnumber=1,
	numbersep=8pt,
	showstringspaces=false,
	breaklines=true,
	frame=lines,
	backgroundcolor=\color{background},
	literate=
	{:}{{{\color{punct}{:}}}}{1}
	{,}{{{\color{punct}{,}}}}{1}
	{\{}{{{\color{delim}{\{}}}}{1}
	{\}}{{{\color{delim}{\}}}}}{1}
	{[}{{{\color{delim}{[}}}}{1}
	{]}{{{\color{delim}{]}}}}{1},
}

\copyrightyear{2021} 
\acmYear{2021} 
\setcopyright{acmlicensed}\acmConference[WSDM '23]{The 16th ACM International WSDM Conference}{February 27, 2023}{Singapore}
\acmBooktitle{ACM International WSDM Conference (WSDM '23), February 27, 2023, Singapore}
\acmPrice{15.00}
\acmDOI{10.1145/3503516.3503530}
\acmISBN{978-1-4503-9599-1/21/12}
\begin{document}
\title{MeSH Suggester: A Library and System for MeSH Term Suggestion for Systematic Review Boolean Query Construction}

%\author{Shuai Wang, Hang Li,Harrisen Scells, Guido Zuccon}
%\affiliation{
%\institution{University of Queensland}
%\city{Brisbane}
%\country{Australia}
%}
%\email{{shuai.wang2, h.li, h.scells, g.zuccon}@uq.edu.au}

\author{Shuai Wang}
\affiliation{%
	\institution{University of Queensland}
	\city{Brisbane}
	\country{Australia}
}
\email{shuai.wang2@uq.edu.au}

\author{Hang Li}
\affiliation{%
  \institution{University of Queensland}
  \city{Brisbane}
  \country{Australia}
}
\email{hang.li@uq.edu.au}

%\author{Bevan Koopman}
%\affiliation{%
%	\institution{CSIRO}
%	\city{Brisbane}
%	\country{Australia}
%}
%\email{bevan.koopman@csiro.au}

\author{Guido Zuccon}
\affiliation{%
  \institution{University of Queensland}
  \city{Brisbane}
  \country{Australia}
}
\email{g.zuccon@uq.edu.au}

\begin{abstract}
	
Boolean query construction is often critical for medical systematic review literature search. To create an effective Boolean query, systematic review researchers typically spend weeks coming up with effective query terms and combinations. One challenge to creating an effective systematic review Boolean query is the selection of effective MeSH Terms to include in the query. In our previous work, we created neural MeSH term suggestion methods and compared them to state-of-the-art MeSH term suggestion methods. We found neural MeSH term suggestion methods to be highly effective. 

In this demonstration, we build upon our previous work by creating (1) a Web-based MeSH term suggestion prototype system that allows users to obtain suggestions from a number of underlying methods and (2) a Python library that implements ours and others' MeSH term suggestion methods and that is aimed at researchers who want to further investigate, create or deploy such type of methods. 
We describe the architecture of the web-based system and how to use it for the MeSH term suggestion task. For the Python library, we 
describe how the library can be used for advancing further research and experimentation, and we validate the results of the methods contained in the library on standard datasets.
Our web-based prototype system is available at \url{http://ielab-mesh-suggest.uqcloud.net}, while our Python library is at \url{https://github.com/ielab/meshsuggestlib}.

\end{abstract}


%\keywords{Systematic Reviews, Query Formulation, Boolean Queries}

\maketitle

\section{Introduction}

Medical systematic reviews are high-quality and comprehensive literature reviews with respect to specific medical research questions. To achieve high effectiveness and efficiency in medical systematic reviews, a high-quality search on medical literature repositories such as PubMed  and Cochrane is the first and most crucial step to gathering enough evidence to support or refute the hypothesis of the review. However, these searches depend strongly on the quality of the search queries~\cite{scells2018generating, wang2022seedcollection}. A high-quality search query may help researchers to gather enough evidence at the minimum cost, as less irrelevant literature will be retrieved. This task is receiving increasing attention from the community~\cite{wang2021mesh, wang2022automated, wang2022seedcollection, scells2017collection, scells2018searchrefiner, lee2018seed, scells2018generating}. The queries used in medical systematic reviews are typically Boolean queries (search terms are combined using 'AND', 'OR' and 'NOT') and often include terms from the Medical Subject Headings (MeSH) \cite{BibEntry2019Sep}. MeSH is a controlled vocabulary thesaurus arranged in a hierarchical tree structure (specificity increases with depth in a parent$\rightarrow$child relationship, e.g., \texttt{Anatomy}$\rightarrow$\texttt{Body Regions}$\rightarrow$\texttt{Head}$\rightarrow$\texttt{Eye}\dots etc.).

However, due to MeSH's large vocabulary size and the systematic review researchers being often unfamiliar with MeSH definitions, selecting suitable MeSH terms to use for a query is challenging. 
Methods for the automatic suggestion of MeSH terms given a query have been devised, with the Automatic Term Mapping (ATM) being currently deployed within PubMed \cite{carlin2004pubmed}. These methods examine a keyword-based query as input (often containing also Boolean operators) and output one or more MeSH terms (sometimes directly in the context of the structured Boolean query). For example for the Boolean query \texttt{TB[tiab] OR tuberculosis[tiab] OR MDR-TB[tiab] OR XDR-TB[tiab]}, ATM suggests the MeSH term \texttt{extensively drug-resistant tuberculosis[MeSH]}.

In this demonstration paper we build upon our previous work on effective methods for MeSH term suggestion~\cite{wang2021mesh, wang2022automated}
 and release a library with associated prototype web system (service and front-end) that implements a number of MeSH term suggestion methods, including ATM and neural methods. We are not aware of any other research that implement methods for the MeSH Term Suggestion task. The library and web service can be integrated into search services that seek to help users creating Boolean queries for medical systematic reviews, e.g. searchRefiner~\cite{scells2018searchrefiner} or PubMed itself. The library can also be used by others wanting to develop new MeSH term suggestion methods as the library is fully extensible and already includes standard evaluation resources (datasets, measures, baselines). The web front-end can be used by researchers wanting to demonstrate their MeSH term suggestion methods, or by users that want to identify the most effective MeSH terms for a query.

%However, with the development of large pre-trained language models, it has been discovered that Boolean queries with MeSH terms suggested by ATM may not be as effective as MeSH Terms suggested by fine-tuned language models in the medical systematic review literature search.
%Therefore, we extend our previous work in MeSH Term suggestion by building a tool and library package for MeSH Term suggestion using fine-tuned language models. From our work, we hope we can not only help suggest better MeSH Terms for systematic review researchers but also help future researchers build up more methods for MeSH Term suggestions easily. 
%
%In this work, we focus on two parts of our implementation.
%
%\begin{enumerate}[leftmargin=*]
%	\item A web-based MeSH Term suggestion tool with accessible API request which supports all BERT-based MeSH Term suggestion method proposed in our previous work.
%	\item An easy to use and integrate MeSH Term suggestion library built for future research on MeSH Term suggestion methods.
%\end{enumerate}

%The problem faced by information specialists is that it is difficult to identify \textit{which} MeSH terms to add to a query.
%In this paper, we are aiming to empirically evaluate a set of MeSH suggestion methods including the most commonly used PubMed ATM method, providing a comprehensive evaluation of MeSH suggestion techniques and proposing new MeSH term suggestion method which adopts the entity retrieval model technique. 

%\begin{description}
%	\item[RQ1] 
%\end{description}

\section{MeSH Term Suggestion Methods}
\label{methods}
Our library currently implements six MeSH Term suggestion methods from two broad families of methods: Lexical (the first three below) and Neural (the remaining): 

%We show six implemented MeSH Term suggestion methods in our web tool and library, including Lexical methods including ATM, MetaMap and UMLS, and Neural methods using BERT, including Atomic-BERT, Semantic-BERT and Fragment-BERT. 

\begin{enumerate}[leftmargin=*]
\item \textbf{ATM} refers to the method currently deployed as part of PubMed for mapping free text into MeSH Terms, journal names or author names. Mapping occurs through the use of rules and mapping tables. We use the ATM implementation available through the PubMed Entrez API \cite{sayers2010general}. 
%\todo{rewrite by concisely saying what ATM does; and yes, include you use the API}
\item \textbf{MetaMap} refers to using the MetaMap tool~\cite{aronson2001effective} to identify medical concepts in queries; the concepts that include entries from the MeSH hierarchy are then used as suggestions.

\item \textbf{UMLS} refers to searching through a purposely built search service we setup based on Elasticsearch v7.6. The index consists of UMLS concepts \cite{bodenreider2004unified}: these include MeSH Terms. The search is performed by issuing the free-text query to the service; the retrieved items are filtered by only including items of the type ``MeSH Terms''; a cut-off may be applied to the resulting ranking.  

\item \textbf{Atomic-BERT} refers to ranking MeSH Terms using the underlying dense retriever to rank MeSH terms with respect to each keyword in the query; we then return the top-ranked MeSH Term.

\item \textbf{Fragment-BERT} refers to performing Atomic-BERT, but before selecting the MeSH Term to suggest, the rankings of the individual query keywords are interpolated using normalised CombSUM rank fusion. The top-ranked MeSH Term is then returned.

\item \textbf{Semantic-BERT} is similar to Fragment-BERT, but the rank fusion is performed with respect to keyword groups rather than across all keywords. Keyword groups are identified based on similarity as computed by a word2vec model trained on PubMed. The top-ranked MeSH term for each Keyword group is then returned as the suggestion.
%	refers to first identifying keywords in a Boolean query fragment to keyword groups, each containing keywords semantically similar to each other. We measure the similarity using w2v models pre-trained on PubMed. We then interpolate the rank of each keyword in a keyword group and select the top one MeSH Term in the interpolated rank as the MeSH Term suggested.
	
\end{enumerate}

All Neural methods use our fine-tuned dual-encoder model described in previous work~\cite{wang2022automated}.

\section{System Overview}
\subsection{MeSH Term Suggestion Web Tool}

\begin{figure}[t!]
	\includegraphics[width=0.85\columnwidth]{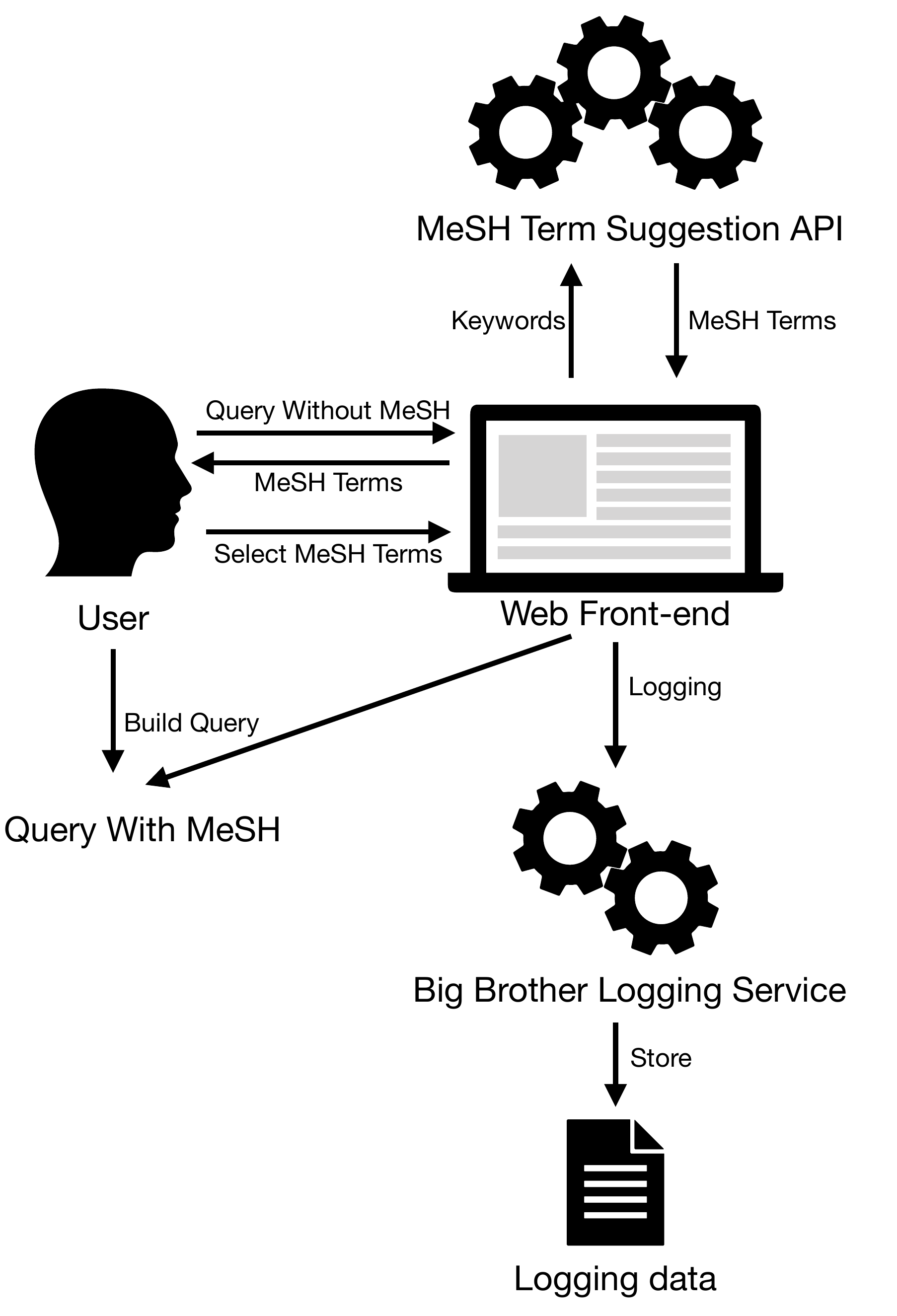}
	\vspace{-10pt}
	\caption{Architecture of web-based MeSH Suggestion tool}
	\vspace{-5pt}
	\label{fig:architecture}
\end{figure}

We start by describing the web service and associated front-end that exposes the implemented MeSH Term suggestion methods. The architecture of the system is provided in  Figure \ref{fig:architecture}. The system consists of (1) the MeSH Term Suggestion API, which wraps the library implementing the suggestions methods, and described in Section~\ref{methods}, (2) the Web front-end, which allows users to enter their keyword queries and receive back the suggestions, (3) the Big Brother logging service, which captures and stores users interactions for subsequent analysis.

Apart from direct usage through the web front-end, we also provide an API for MeSH Term suggestions. 

\begin{figure}[t!]
	\begin{lstlisting}[language=json,firstnumber=1]
{"Keywords": [K1, K2, K3, ..., Kn],
	"Type": Semantic/Atomic/Fragment}
	\end{lstlisting}
	\vspace{-10pt}
	\caption{Input format of the API POST call.}
	\vspace{-10pt}
	\label{fig:input-api}
\end{figure}

\begin{figure}[t!]
	\begin{lstlisting}[language=json,firstnumber=1]
[	
{"Keywords": [K1],
	"Type": Semantic/Atomic/Fragment},
"MeSH_Terms": {0: M1, 1: M2, ...., 9: M10}

{"Keywords": [K2, K3],
	"Type": Semantic/Atomic/Fragment},
"MeSH_Terms": {0: M1, 1: M2, ...., 9: M10}
...
{"Keywords": [Km, ... Kn],
	"Type": Semantic/Atomic/Fragment},
"MeSH_Terms": {0: M1, 1: M2, ...., 9: M10}
]
	\end{lstlisting}
	\vspace{-10pt}
	\caption{Output format from the API POST call.}
	\vspace{-10pt}
	\label{fig:output-api}
\end{figure}

The MeSH Term suggestion API exposes to users the POST method to call the API that, provided a query, returns a list of MeSH Term suggestions using one of the implemented methods. The input format is shown in Figure \ref{fig:input-api}, while the output of the call is shown in Figure \ref{fig:output-api}. The API output includes the original keyword query input,  the suggestion type (i.e. the method used to generate the suggestion), and the MeSH Terms suggested for each keyword or keywords group.

\begin{figure}[t!]
	\includegraphics[width=0.85\columnwidth]{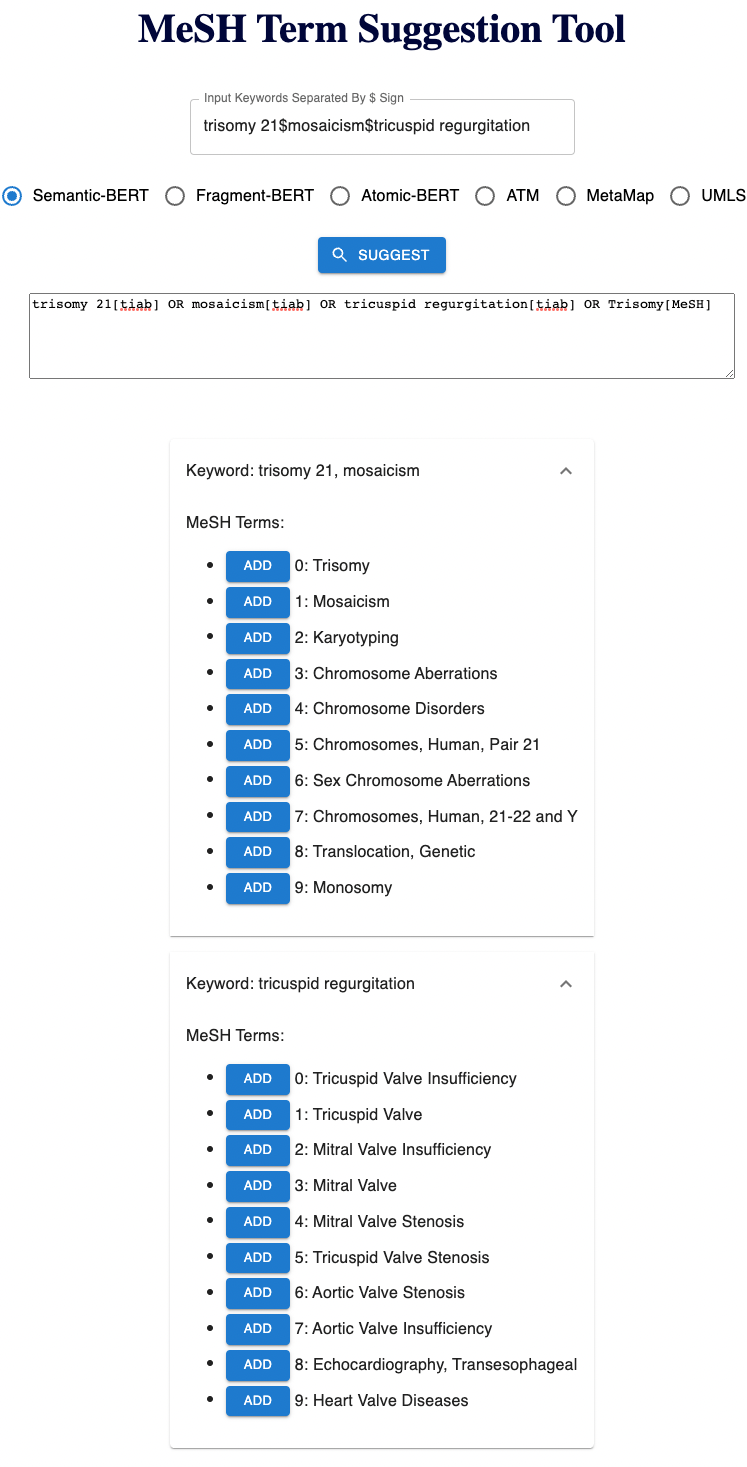}
		\caption{Example Mesh Term suggestion using our Web tool.}
		\vspace{-15pt}
	\label{fig:web-tool-suggested}
\end{figure}

The web frond-end is shown in Figure \ref{fig:web-tool-suggested}. Users can submit a single keyword or keywords combination and choose to use any of the methods outlined in Section~\ref{methods}. Upon submission of a query, the tool returns a list of candidate MeSH Terms that the user can copy or use the inbuilt tool to add to their free-text query to form a new query with MeSH terms (which they can eventually copy). Currently, the Boolean query construction box is naively appending newly added terms by "OR" as we do not identify this as a target task for this paper. Future works on how to lead users to issue more effective queries will also be investigated to help for a more effective Boolean Query generation for systematic review literature search.

The interaction logging service, Big Brother \cite{10.1145/3404835.3462781}, is also integrated into our tool's front-end and captures all interactions of the users with the web page. The logging service may help with the future investigation of MeSH Term suggestion methods through user studies.

%When the user type in their keyword, they can receive MeSH Term suggestions as ranked lists with respect to a specific keyword or identified keywords group, as shown in Figure \ref{fig:web-tool-suggested}. The user then may click the MeSH Term to add to construct a Boolean query. Our web-based tool also embeds a drop-in website interaction logging service; namely Big Brother \cite{10.1145/3404835.3462781}. The logging service can help with future user studies on MeSH Term suggestion or evaluation of MeSH Term suggestion methods. The architecture of our web-based tool is shown in Figure \ref{fig:architecture}.

\subsection{MeSH Term Suggestion Library}
Along with the web service API and web front-end described above, we also provide a Python-based library package, \texttt{meshsuggestlib}, that implements the methods described in Section~\ref{methods}. The package also makes available classes that can be extended for the implementation of new MeSH Term suggestion methods. Finally, the package includes data and associated auxiliary code for evaluating MeSH Term suggestion methods. These inclusions allow others to quickly implement, validate and compare new MeSH Term suggestion methods. For example, the results for the Semantic-BERT MeSH Term suggestion methods on the CLEF TAR 2017 dataset~\cite{kanoulas2017clef}, which we show in Table~\ref{table:retrieve}, can be obtained by running the following commands:

%In addition, we build a Python package called `meshsuggestlib' for future research on the MeSH Term suggestion task. Our tool follows the structure of popular Dense retriever tool such as Tevatron \cite{Gao2022TevatronAE} and Asyncval \cite{10.1145/3477495.3531658}.

%In our package, we set up pre-built dataset and methods for researchers who wish to reproduce our previous experiment. For example, running the following command outputs Semantic-BERT MeSH Term suggestion results using CLEF Tar 2017 dataset \cite{kanoulas2017clef}:
\begin{lstlisting}[language=bash]
	python -m meshsuggestlib
	--model_dir model/checkpoint-80000/
	--method Semantic-BERT
	--dataset CLEF-2017
	--output_file out.tsv
	--email sample@gmail.com
	--interpolation_depth 20
	--depth 1
\end{lstlisting}

Similarly, these results can be evaluated with simplicity using the following template command:

\begin{lstlisting}[language=bash]
	python -m meshsuggestlib
	--evaluate_run
	--output_file out.tsv
	--qrel qrel.qrels
\end{lstlisting}
 
Table~\ref{table:input} reports a full list of input options for \texttt{meshsuggestlib}. For the neural models we implemented, it is possible to change the underlying model checkpoint used, although currently only dense retrievers (bi-encoders) are supported. Nevertheless, it is possible for researchers to extend the package by implementing new MeSH Term suggestion methods, or adding new evaluation datasets; we show how one can add a new suggestion method in Section~\ref{addnewmethod}.

\begin{table}[t!]
	\centering
	\scriptsize
	\begin{tabular}{p{0.1pt}ll}
		\toprule
		&Input Name & Description \\
		\midrule
		\multirow{3}{*}{\rotatebox{90}{Basic}}&$method$	& Predefined MeSH Term Suggestion method or new method\\
		&$dataset$ & Pre-defined Dataset or data folder name \\
		&$mesh\_file$ & MeSH Term file path \\ \midrule
		\multirow{5}{*}{\rotatebox{90}{Neural}}&$mesh\_encoding$ & [Optional] path of encoded MeSH Terms \\
		&$tokenizer\_name\_or\_path$ & Tokenizer for Neural Methods \\
		&$model\_dir$ & Neural Model path or name \\
		&$q\_max\_len$ & query keyword maximum length after tokenization \\
		&$p\_max\_len$ & MeSH Term maximum length after tokenization\\ \midrule
		\multirow{3}{*}{\rotatebox{90}{Group}}&$semantic\_model\_path$ & Path of w2v Model for semantic grouping\\
		&$interpolation\_depth$ & Cut-off of each keyword for interpolation\\
		&$depth$ & Cut-off for number of MeSH Term retrieved for each group\\\midrule
		\multirow{3}{*}{\rotatebox{90}{PubMed}}&$output\_file$ & Path of query result output\\
		&$date\_file$ & Path of date restriction file for each topic\\
		&$email$ & Email for calling E-utilities API for literature retrieval\\\midrule
		\multirow{2}{*}{\rotatebox{90}{Evaluate}}&$evaluate\_run$ & Whether evaluate the output result \\
		\\
		&$qrel\_file$ & Path to file containing relevance judgments\\
		\bottomrule
	\end{tabular}
	\vspace{-15pt}
	\caption{Input option for library.\vspace{-20pt}}
	\label{table:input}
\end{table}

%We show a full list of input options for researchers to use in table \ref{table:input}. Researchers can also run experiments using a different fine-tuned model they train; currently, only Dense retrievers are supported in our MeSH Term suggestion methods. Alternatively, researchers may use other systematic review dataset for MeSH Term suggestion or create new search function for the MeSH Term suggestion task.

\vspace{-5pt}
\section{Case Studies}

Next, we report on a small-scale validation of the methods we implement in \texttt{meshsuggestlib} and the associated web tool; then, we describe how the library can be expanded by implementing new MeSH Term suggestion methods. 
\begin{table}
	\centering
	\small
	\begin{tabular}{p{0.1pt}c|p{17pt}p{17pt}p{18pt}|p{17pt}p{17pt}p{18pt}}
		 \toprule
		\multicolumn{2}{c|}{Dataset}&\multicolumn{3}{c|}{2017}&\multicolumn{3}{c}{2018}\\ \midrule
		\multicolumn{2}{c|}{Method}&\multicolumn{1}{c}{P}&\multicolumn{1}{c}{F1}&\multicolumn{1}{c|}{R}&\multicolumn{1}{c}{P}&\multicolumn{1}{c}{F1}&\multicolumn{1}{c}{R}\\ \midrule
		&Original&0.0303&0.0323&0.7694&0.0226&0.0415&\textbf{0.8629} \\\midrule
		%&Removed&0.0418&0.0353&0.6931&0.0431&0.0762&0.7790 \\\midrule
		\multirow{3}{*}{\rotatebox{90}{Lexical}}&ATM&0.0225&0.0215&0.7109&0.0306&0.0535&0.8225 \\
		&MetaMap& 0.0323&0.0304&0.7487&0.0335&0.0590&0.8085\\ 
		&UMLS&0.0325&0.0300&0.7379&0.0326&0.0573&0.7937 \\ \midrule
		\multirow{3}{*}{\rotatebox{90}{Neural}}& Atomic-BERT&0.0252&0.0243&0.7778&0.0283&0.0479&0.8452\\
		&Semantic-BERT&0.0254&0.0243&\textbf{0.7784}&0.0309&0.0526&0.8404 \\ 
		&Fragment-BERT&\textbf{0.0343}&\textbf{0.0325}&0.7414&\textbf{0.0388}&\textbf{0.0690}&0.8034 \\ 
		\bottomrule
	\end{tabular}
	\caption{Effectiveness of MeSH Term suggestion methods in terms of precision(P), F1 and recall (R). No statistical significance is detected between the Original query and those obtained by other methods (two-tailed t-test with Bonferroni correction, p < 0.05). \vspace{-20pt}}
	\label{table:retrieve}
	\vspace{-5pt}
\end{table}
\vspace{-5pt}
\subsection{Evaluation of Methods}
We evaluate all implemented methods on the CLEF Tar 2017~\cite{kanoulas2017clef} and 2018~ \cite{kanoulas2018clef} datasets.
For each topic in the dataset, we stripped the original Boolean query of the Boolean operators and the MeSH terms, so to obtain a keyword query which was then used as input for the MeSH Term suggestion methods. We then attach the suggested MeSH Terms to the query and use this to retrieve documents from the PubMed index. Evaluation is performed with respect to how effective the query was for retrieval -- the better the query, the more effective the MeSH Term suggestion method. Note, this is a retrieval task, not a ranking task, as queries and the underlying retrieval system are Boolean. Also note that the original query is likely to outperform the automatic queries: this is because these queries have undergone careful manual intervention by information specialists. We refer to our previous work for more details of the evaluation setup~\cite{wang2022automated}.

 Results are reported in Table \ref{table:retrieve}, where we also include the results obtained on the original Boolean query (which includes MeSH terms added by information specialists).

%on six suggestion methods using our package, including Lexical methods: ATM, MetaMap and UMLS, and Neural Methods: Atomic-BERT, Semantic-BERT and Fragment-BERT. 
%We also report the original Boolean query from the CLEF-Tar dataset issued on PubMed. We show our result in Table \ref{table:retrieve}. Note that the results are slightly different from the results from our previous paper \cite{wang2022automated}; this is due to several factors: 

Results differ from our recent evaluation of these methods (see~\cite{wang2022automated}) because:
(1) We issue our constructed query to PubMed's E-Utilities API \cite{sayers2010general} to retrieve documents for evaluation; some PubMed articles may be changed or updated; thus may be filtered out by the Boolean keywords or date restrictions. 
(2) For the Lexical methods, we use the PubMed API for ATM, and the UMLS and Metamap for the other methods; the implementations and the data used by these methods may have received updates between the two undertaking of the experiments.
For Neural methods, the encoder integrated in the library has been retrained and thus may differ from that originally used in previous work because of small differences in initial weights and training process.
(3) The evaluation is conducted using the \texttt{ir\_measures} toolkit~\cite{macavaney2022streamlining} instead of \texttt{Trec\_eval} because of its better fit into our Python library --these two tools have minor differences in how recall is computed. 
Despite these aspects, the trend we observe from the results is the same, and the differences between our reproduced results and previous experiments are marginal.

%Additionally, it is possible to use a new dataset to evaluate all MeSH Term suggestion methods using our package, as long as the dataset includes Boolean queries and relevance judgement of retrieved documents. The new dataset should follow the same structure as our processed CLEF Tar dataset in package. Specifically, the Boolean query of each topic should be dissembled to query fragments as in our previous paper \cite{wang2021mesh}. Each query fragment should be inside a new folder under the topic folder. Detailed instructions on using new dataset can be found in our Github repository.

%For our MeSH Term suggestion web tool, we would like to include systematic reviewers as our target user group. From our previous works, we believe MeSH Terms suggested by our tool may bring more effective MeSH Term suggestions than the current state-of-the-art methods such as ATM, thus may bring more effective MeSH Term selection for our user group. However, we evaluated our MeSH Term suggestion methods using the effectiveness of retrieved documents from the MeSH Term suggested query and original query using CLEF TAR datasets. The evaluation may be biased as a selection of documents from MeSH Term suggested queries were never judged. With the help of a practical MeSH Term suggestion system and users in the systematic review community, we may better understand how to evaluate the effectiveness of MeSH Term suggestion methods in systematic review Boolean query construction.

\vspace{-7pt}
\subsection{Add a new MeSH Term Suggestion Method}
\label{addnewmethod}
\texttt{meshsuggestlib} allows researchers to implement new suggestion methods. If these methods are based on the neural architecture used by our methods, then it is sufficient to change the library input parameters \textit{tokenizer\_name\_or\_path} and  \textit{model\_dir} to direct the library to the new dense retriever models. If instead the underlying retrieval logic differs, to add a new method it is sufficient to implement the search function  \textit{user\_defined\_method} in the \textit{NeuralSuggest} class.  This function takes keywords, the retriever models and lookups as input and returns a list of keywords and MeSH Term IDs pairs as output. At inference, the use of the method `NEW' will automatically call this function.

%Researchers can also build a new MeSH Term suggestion method and integrate it with our library to evaluate. If the new method follows the same retrieval pipeline with only a different Dense retriever model, researchers can only change input parameters \textit{tokenizer\_name\_or\_path} and  \textit{model\_dir} to test their new models. Alternatively, if the underlying retrieval logic differs, researchers only need to write the search function to run the full pipeline. We reserved a function called \textit{user\_defined\_method}  in our \textit{NeuralSuggest} class. This function takes keywords, retriever and lookups as input and returns a list of keywords and MeSH Term IDs pairs as output. During inference, using method `NEW' will automatically call this function for the retrieval task.

\vspace{-5pt}
%\subsection{Package for MeSH Term Suggestion Research}

%The MeSH Term suggestion package is specifically built for researchers who wish to help with more effective MeSH Term suggestions during Boolean query creation in the systematic review. Not only can users easily reproduce results from our proposed MeSH Term suggestion methods, but also easily integrate new methods, keyword grouping strategies, and cut-off strategies. 
% Furthermore, users may also evaluate MeSH Term suggestion methods using our pre-built evaluation metrics and datasets; thus, reporting MeSH Term suggestion results can be fair and straightforward.
%\input{discussion}
\section{Conclusion \& Discussion}
This demonstration contributes useful tools for the MeSH Term suggestion task: a library that implements common lexical baselines and neural methods, and a web service with associated web front-end that allows end-users to use these methods to augment their queries for systematic review literature search. The tool also allows to collect usage and interaction logs, thus allowing researchers to further their understanding of MeSH Term choices and the query formulation process~\cite{scells2022impact}. The library also integrates an evaluation pipeline, including the implementation of accessory methods for standard datasets in this context: this lowers the barrier for others to research new MeSH Term suggestion methods. 

Several improvements are currently planned for the tool. A key feature to further streamline use of the tool is the automatic decomposition of Boolean queries and the related extraction of keywords, which are then used as input to the MeSH Term suggestion methods. Another avenue of improvement is integrating the library into existing Boolean query visualisation tools, like  SearchRefiner~\cite{scells2018searchrefiner}, which allows users to interpret how the choices made with respect to the MeSH terms suggested affect retrieval and effectiveness. 

\subsubsection*{Acknowledgement} 
\small This research is supported by the Australian Research Council (DP210104043).
\small The authors of the work also wish to thank Dr Harrisen Scells for providing instruction and suggestions.

\bibliographystyle{ACM-Reference-Format}
\bibliography{references.bib}

\end{document}